\documentclass{jltp}
\usepackage{graphicx}

\title{ Decay of turbulence generated by spin-down to rest in superfluid $^4$He }

\author{P.\,M.\,Walmsley$^a$, A.\,I.\,Golov$^a$, H.\,E.\,Hall$^a$, \\ W.\,F.\,Vinen$^b$, and A.\,A.\,Levchenko$^c$}

\address{$^a$School of Physics and Astronomy, The University of Manchester, M13 9PL, UK \\
$^b$School of Physics and Astronomy, University of Birmingham, B15 2TT, UK \\
$^c$Institute of Solid State Physics, Russ. Acad. Sci., Chernogolovka 142432, Russia}

\runninghead{P. M. Walmsley {\it et al.}}{Turbulence in superfluid $^4$He generated by spin-down}

\begin{document}

\maketitle

\begin{abstract}
We report on the extension of the experiments (P.\,M.\,Walmsley {\it et al.}, Phys. Rev. Lett. {\bf 99}, 265302 (2007)) on the decay of quasiclassical turbulence generated by an impulsive spin-down from angular velocity $\Omega$ to rest of superfluid $^4$He in a cubic container at temperatures 0.15\,K -- 1.6\,K. The density of quantized vortex lines $L$ is measured by scattering negative ions. Following the spin-down, the maximal density of vortices is observed after time $t \sim 10\Omega^{-1}$. By observing the propagation of ions along the axis of the initial rotation, the transient dynamics of the turbulence spreading from the perimeter of the container into its central region is investigated. Nearly homogeneous turbulence develops after time $t \sim 100\Omega^{-1}$ and decays as $L \propto t^{-3/2}$. The effective kinematic viscosity in $T=0$ limit is $\nu = 0.003 \kappa$, where $\kappa=10^{-3}$~cm$^2$s$^{-1}$ is the circulation quantum. 

PACS numbers: 67.25.dk, 47.27.Gs, 47.32.-y, 47.37.+q.
\end{abstract}
\section{INTRODUCTION}

Turbulence in superfluid helium \cite{VinenNiemela2002,Vinen2006,EltsovPLTP2008} in the $T=0$ limit can be fully characterized by the positions and dynamics of thin cores of quantized vortices making a dense tangle. A useful  parameter is the total length of these filaments per unit volume $L$, the mean inter-vortex distance hence being $\ell = L^{-1/2}$. Different spatial distributions of local density $L(\bf{r})$ and local polarization of the tangle correspond to different types of turbulent flows. When turbulence is generated by flow on large (quasiclassical) lengthscales $\gg \ell$, its dynamics is of great interest as the dissipative mechanisms only work at much smaller lengthscales. The energy is injected into large
quasi-classical eddies which are a result of correlations in the polarization of vortex lines. While cascading down the lengthscales, this classical energy should be eventually converted into that of quantized flow around individual vortices. The latter have the following degrees of freedom: deformation of the individual filaments (Kelvin waves) and reconnections of pairs of neigbouring filaments (change in the tangle's topology). The ultimate mechanism of the dissipation of energy is believed to be phonon emission by Kelvin waves of very short wavelength. Vortex reconnections leave sharp kinks, thus exciting Kelvin waves. One of the modern challenges is to understand the nature of the non-linear cascades of energy down the classical ($> \ell$) and quantum ($< \ell$) lengthscales, and especially of its transfer from the classical to quantum cascade at the cross-over scale $\sim \ell$. 

Very recently \cite{Walmsley2007} there has been progress in generating quasi-classical turbulence by spin-down of a non-axially-symmetric container to rest and detecting the decay of the vortex density $L(t)$ through scattering of injected ions off vortex filaments. The latter allowed us to quantify the rate of the energy dissipation at different temperatures. For the given $L$, this rate can be different for different types of turbulence \cite{Walmsley2008}, especially in the $T=0$ limit. It is believed that the quasi-classical turbulence achieved in the mentioned experiments is nearly homogeneous and isotropic. In this paper we provide further details on the properties of the turbulent tangles obtained in this way, discuss the appropriateness of the analysis, introduce a refined fitting procedure and summarize the current understanding of the nature of the energy cascade and dissipation. 

The assumptions for the analysis of the late-time decay $L(t)$ are as follows \cite{Stalp1999}. On the classical side of wavelengths, the energy spectrum is of the Kolmogorov type
\begin{equation}
E_k = C \epsilon^{2/3} k^{-5/3},
 \label{Ek}
\end{equation}
where the Kolmogorov constant for classical fluids was found to be $C\approx 1.5$
\cite{Sreeni1995}. Because of the confinement by the container of size $d$, this spectrum is truncated at short wavenumbers by $k_1 \approx 2\pi/d$ making $d$ the energy-containing length. 

At short length scales $<\ell$ the
cascade becomes {\it quantum} as the discreteness of the vorticity
in superfluids adds new behavior. At sufficiently low temperatures, it is expected to take the
form of a non-linear cascade of Kelvin waves
\cite{S1995,KS2004}. The most complicated regime is the transitional region between these two 
limits, the Kolmogorov and Kelvin-wave cascades. The question
being currently debated is whether the energy stagnates at $k <
\ell^{-1}$ due to the poor matching in the kinetic times of the
two cascades \cite{LNR2007}, or gets efficiently converted from 
classical eddies to waves along quantized vortex lines with
the help of various reconnection processes \cite{KS2008}.

In the steady state, the energy flux $\epsilon$ is
equal to the dissipation of the kinetic energy. 
  The redistribution between different lengthscales and eventual dissipation of the flow energy is through the motion of vortex
lines. The flux down the classical cascade of this energy per unit mass can be assumed to be
\cite{Stalp1999}
\begin{equation}
    \epsilon = \nu(\kappa L)^2,
    \label{EDot}
\end{equation}
where $\kappa^2 L^2$ is an effective total mean square vorticity
and the ``effective kinematic viscosity'' $\nu$ can vary depending upon temperature and type of flow. With regard to the physical meaning of Eq.\,(\ref{EDot}) there are two possibilities. It can be simply an analog of the classical formula for the viscous dissipation rate in shear flow at the smallest dissipative lengths. Alternatively, it can result from a cascade-specific process (like ``bottleneck'') limiting the energy flux $\epsilon$ down the lengthscales.  

Equations \ref{Ek} and \ref{EDot} lead to the
late-time free decay  \cite{Stalp1999}
\begin{equation}
    L = (3C)^{3/2}\kappa^{-1}k_1^{-1}\nu^{-1/2} t^{-3/2}.
    \label{L-t-K}
\end{equation}
In the rest of this paper we show that this regime is observed for quasi-classical turbulence in superfluid $^4$He generated by various means, including the $T=0$ limit \cite{Stalp1999,Walmsley2007,Walmsley2008}. It is also observed for vortex tangles in superfluid $^3$He-B generated by a vibrating grid \cite{Bradley2006}. 
 
\section{Experimental}

To detect the presence of vortices we used the fact that injected ions interact with a vortex. This process is characterized by the trapping (or ``scattering'') diameter $\sigma$ which depends on temperature, electric field and type of ion. Two types of charge carriers have been used: at $T>0.8$\,K, free ions (an electron self-localized in a bubble), and at $T<0.8$\,K small charged vortex rings (CVRs) with one negative ion trapped on their cores. 

For free ions, the trapping diameter $\sigma$ is inversely proportional to the electric field, $\sigma
\propto E^{-1}$. In a field $E=10$\,V/cm it is decreasing with
temperature from $\sigma = 220$\,nm at $T=1.6$\,K to $\sigma =
4$\,nm at $T=0.8$\,K. In contrast, CVRs have a much larger trapping diameter of order their radius (typically, $\sigma = 0.4$--$1.7$\,$\mu$m in our experiments) which makes them much more sensitive to tangles of small density $L$. 

\begin{figure}[h]
\begin{center}
\includegraphics[width=0.75\textwidth]{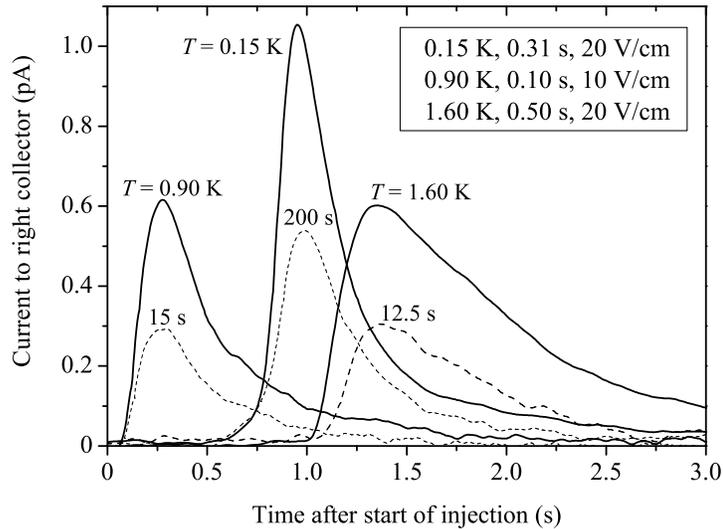}
\caption{Examples of current records produced by short pulses
for three different temperatures (the temperatures, pulse
durations and mean driving fields are indicated in the legend).
The solid lines show the records without a vortex tangle in the
ions' path, while the dashed ones represent the records
suppressed by the vortex  tangle which has been decaying a
specified time (indicated near curves) after stopping generation.
The charge carriers are either free ions ($T=1.60$\,K and
$T=0.90$\,K) or charged vortex rings ($T=0.15$\,K). The electronics
time constant is 0.15\,s, hence the true time of arrival of the ions producing the fastest
peak ($T=0.90$\,K) cannot be resolved.}
 \label{MancPulseExamples}
\end{center}
\end{figure}

\begin{figure}[h]
\begin{center}
\includegraphics[width=0.75\textwidth]{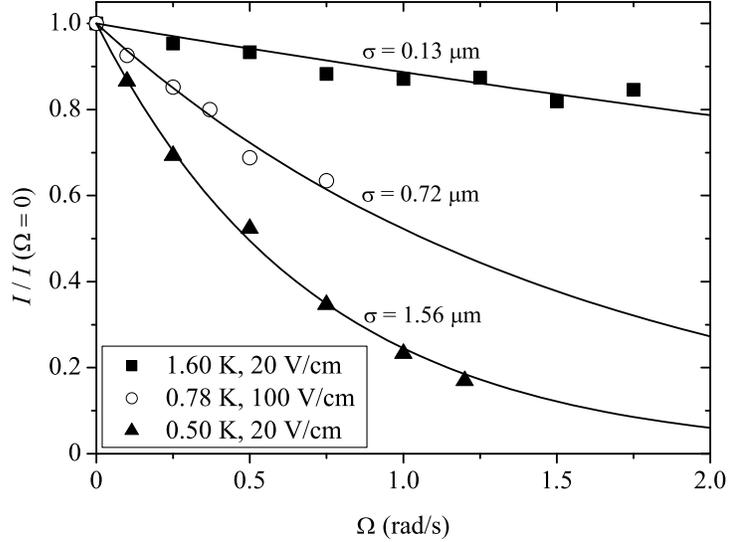}
\caption{Dependence of pulse amplitude on the angular velocity of
rotation, $\Omega$, in examples of calibration measurements. The
temperatures, driving fields and trapping diameters $\sigma$ are
indicated. The charge carriers are either free ions ($T=1.60$\,K)
or charged vortex rings ($T=0.78$\,K and $T=0.50$\,K). }
 \label{MancCrossSections}
\end{center}
\end{figure}

After
injecting a short pulse of ions, a sharp pulse of current
arrives at the collector as shown in Fig.\,\ref{MancPulseExamples}.
The interaction with vortices, that happend to be
in the way of the propagating ions leads to the depletion of these
pulses. The decrease in the peak height $I$ of the current pulses due to the ballistic CVRs (shown in Fig.\,\ref{MancPulseExamples}), relative to the long time amplitude $I(\infty)$ when all vortices have decayed, is the measure of blocking the paths of some ions with the probability per unit length $L \sigma$. We hence determine the average vortex density between the injector and collector, after the turbulence has been decaying a time $t$, as
\begin{equation}
L(t) = (\sigma d)^{-1}\ln\left(\frac{I(\infty)}{I(t)}\right).
\end{equation}
To calibrate the value of $\sigma$ for different types of ions and at different temperature and
electric field, we measured the attenuation of the pulses of
the collector current in the horizontal direction when the cryostat is at continuous rotation \cite{MancRotCryostat1,MancRotCryostat2} around the vertical axis 
at angular velocity $\Omega$ (thus having an equilibrium density
of rectilinear vortex lines $L=2\Omega/\kappa$) as in Fig.\,\ref{MancCrossSections}. 

At low temperatures, Kelvin waves of a
broad range of wavelengths are excited, thus making the tangle length a fractal property. 
However, the main
contribution to the total length $L$ converges quickly at scales
just below $\ell$. Hence, a probe ion with
the trapping diameter $\sigma \ll \ell$, moving at a speed ($v
\geq 5$\,cm/s), much greater than the characteristic velocities of
vortex segments ($\sim \kappa/\ell < 10^{-1}$\,cm/s), should sample
the full length $L$. Experimental measurements with different
$\sigma = 0.4$--1.7\,$\mu$m (using charged vortex rings in a range
of driving electric fields) indeed produce consistent values of
$L$.

In order not to introduce extra turbulence by the probe ions, the measurement
was performed by probing each realization of turbulence only once
-- after a particular waiting time $t$ during its free decay. Then
the contaminated tangle was discarded and a fresh tangle
generated. The values of $L$, measured with the use of either of the two types of ions in the temperature window 0.7--0.8\,K where they co-exist, were consistent too. 


\begin{figure}[h]
\begin{center}
\includegraphics[width=0.75\textwidth]{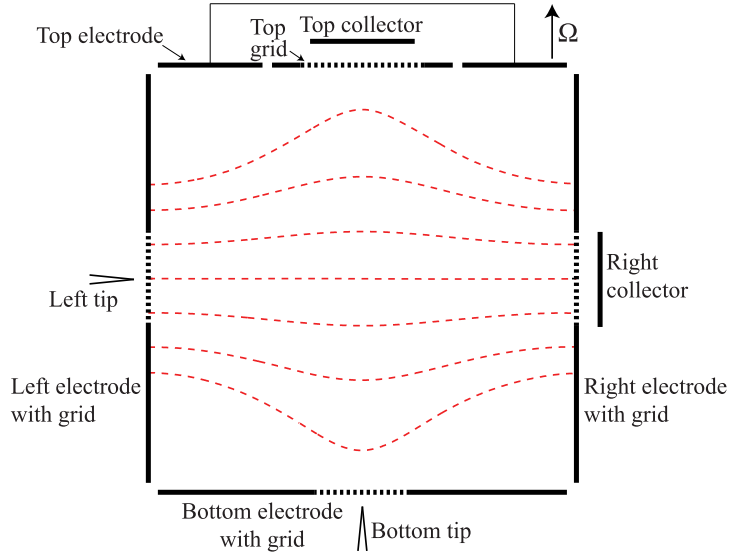}
\caption{A side cross-section of the Manchester cell. The distance
between opposite electrode plates is 4.5\,cm. An example of the
driving electric field is shown by dashed lines. Such a
configuration was used in all the Manchester experiments described
in this section. It was calculated for the following potentials
relative to the right electrode: left electrode at -90\,V, side,
top and bottom electrodes at -45\,V, and the right electrode and
collector at 0 producing a 20 V/cm average driving field in the
horizontal direction. To inject ions, the left tip was usually
kept at  between -500 to -350 V relative to the left grid. When
ions traveling vertically across the cell (injected from bottom
tip and detected at the top collector) were required, the
potentials on the electrodes were rearranged as appropriate. }
\label{MancCellSchematic}
\end{center}
\end{figure}

The experimental cell had cubic geometry
with sides of length $d=4.5$\,cm. A schematic drawing of the cell
is shown in Fig.\,\ref{MancCellSchematic}. The relatively large
size of the cell was important to enhance the efficiency of ion
trapping and the time resolution of vortex dynamics, and was
really instrumental in ensuring that the continuum limit $\ell \ll
d$ holds even when the vortex line density drops to as low as just
$L \sim 10$\,cm$^{-2}$. This also helped ensure that the presence
of the walls, which might accelerate the decay of turbulence
within some distance $\sim \ell$, does not affect the dynamics of
the turbulent tangle in the bulk of the cell. In order to probe
the vortex densities along the axial and transverse directions,
there were two independent pairs of injectors and collectors of
electrons. All injectors and collectors were protected by
electrostatic grids, enabling injecting and detecting pulses of
electrons. The injectors were field emission tips made of 0.1\,mm
diameter tungsten wire \cite{GolovIshimoto1998}.  The threshold for
ion emission was initially $\simeq$ -100\,V and -210 V for the
bottom and left injectors respectively; however, after over two
years of almost daily operation they changed to some -270\,V and
-520\,V. The fact that the two injectors had very different
threshold voltages helped investigate the dependence (or rather
lack of it) of the radius of initial charged vortex rings on the
injector voltage.
 The six side plates
(electrodes) that make up the container cube can be labeled ``top'',
``bottom'', ``left'', ``right'', and two ``side'' electrodes. The
top, bottom, left and right electrodes had circular grids in their
centers. All grids were made of square tungsten mesh with period
0.5\,mm and wire diameter 0.020\,mm, giving a geometrical
transparency of 92\%. The grids in the bottom, left and right
plates had diameter 10\,mm and the grid in the top plate had diameter of 13\,mm. The injector tips were positioned
about 1.5--2\,mm behind their grids. The two collector plates were
placed 2.5\,mm behind their grids and were typically biased at +10
-- +25\,V relative to the grids. Further details can be found in
\cite{Walmsley2006}.


The novel technique of generating quasiclassical turbulence,
suitable for any temperatures down to at least 80\,mK, relied on
rapidly bringing a rotating cubic-shaped container of superfluid
$^4$He to rest \cite{Walmsley2007}. This was achieved by mounting the experiment on the Manchester rotating cryostat \cite{MancRotCryostat1,MancRotCryostat2}, whose angular velocity and acceleration can be controlled over a wide range. In our experiments, the range of angular
velocities of initial rotation $\Omega$ was 0.05--1.5\,rad/s. In
classical liquids at sufficiently high Reynolds numbers, even in cylindrical containers, spin-down to rest is always unstable: within a few radians of initial
rotation upon an impulsive spin-down to rest, a nearly homogeneous
turbulence develops with the energy-containing eddies of the size
of the container. In a cubic container, the
turbulence becomes homogeneous much faster. 
One expects similar processes to occur
in a superfluid liquid providing the process of initial
multiplication of vortices does not affect the dynamics. As
spin-down experiments always begin from already existing dense
rectilinear vortex arrays of equilibrium density
$L=2\Omega/\kappa$, and rapid randomization and multiplication of
these vortices is expected due to the lack of axial symmetry of
the container, as well as surface pinning (friction) near container walls, these seem to be sufficient for the superfluid to
mimic the large classical turbulent eddies. 

Before making each measurement, the cryostat was kept at steady
rotation at the required $\Omega$ for at least 300\,s  before
decelerating to full stop, then waiting a time interval $t$ and
taking the data point. The deceleration was
linear in time taking 2.5\,s for $\Omega = 1.5$\,rad/s and 0.1\,s for
$\Omega = 0.05$\,rad/s. The origin $t=0$ was chosen at the start of
deceleration.

We have also found that currents of injected ions leave vortex tangles behind that can also be used as an alternative means of generating turbulence. 
Depending on the conditions of injection, such as the temperature and duration of the current pulse, the tangles can have features of either quasi-classical (large-scale dominated) or quantum (quantum energy dominated) turbulences. We do not discuss these experiments in this paper but quote the resulting values of $\nu(T)$ in later sections. For more details see \cite{Walmsley2008}. 

\section{Results}

\begin{figure}[h]
\begin{center}
\includegraphics[width=0.75\textwidth]{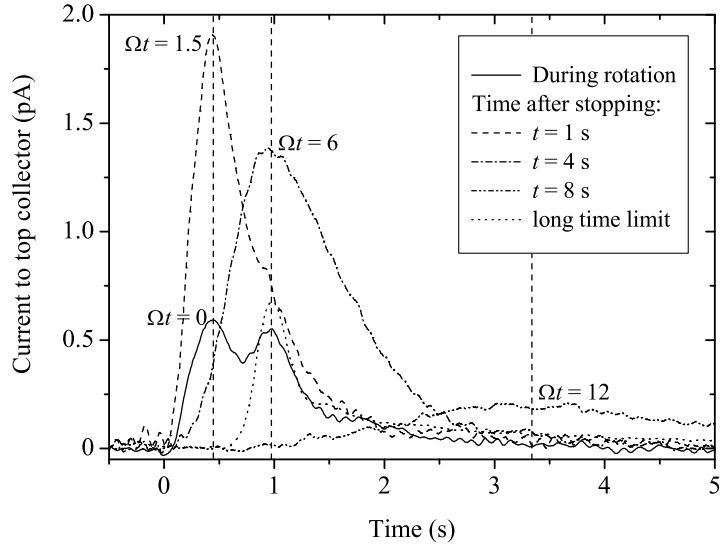}
\caption{Records of the current to the top collector, injected
from the bottom injector as a 0.1\,s-long pulse at the time $t$
after an impulsive spin-down from $\Omega = 1.5$\,rad/s to rest. $T
=0.15$\,K, $E= 20$\,V/cm.}
\label{MancSpinDownTransients}
\end{center}
\end{figure}

To illustrate what is happening near the vertical axis of the
container at different stages of the transient after a spin-down,
in Fig.\,\ref{MancSpinDownTransients} we show five records of the
current from the bottom injector to the top collector arriving
after a 0.1\,s-long pulse of probe ions was fired from the
bottom injector. Each is characteristic of a particular
configuration of vortex lines near the rotational vertical axis of
the container during the transformation from an array of parallel
lines to a homogeneous decaying vortex tangle. One can see three
different characteristic times (vertical dashed lines in
Fig.\,\ref{MancSpinDownTransients}) of arrival of ions via
different means. The first peak at $\approx 0.4$\,s (determined by
the time constant of the current preamplifier, 0.15\,s) corresponds
to the ions trapped on the rectilinear vortex lines which can
slide along those lines very quickly, provided these lines are
continuous from the injector to collector as, for example, during steady
rotation. Our estimates show that at $\Omega =1.5$\,rad/s of order half 
of rectilinear vortices from the equilibrium array terminate on the collector 
grid and hence do not allow trapped ions to reach the collector. 
 The second peak at $\approx 1$\,s corresponds to the
coherent arrival of the ballistic charged vortex rings from the
bottom to the top. The third broad peak at times $\sim 3$\,s (but
with a long tail detectable until $\sim 40$\,s) corresponds to the
charge trapped on the vortex tangle and drifting with the tangle.
Hence, we have the following regimes (and curves in
Fig.\,\ref{MancSpinDownTransients}) labeled by the time $t$ after
the spin-down:
\begin{enumerate} 
\item{$\Omega t\leq 0$, steady
rotation. The nearly equal first and second peaks indicate that there
exist an ordered array as the ions can arrive at the collector by either fast sliding along the rectilinear lines or propagating as slower CVRs between the lines.}
\item{$\Omega t = 1.5$. The first peak gets enhanced three-fold
while the second one is still there (and no third peak) -- meaning
more trapped ions can now reach the collector along the
rectilinear vortex lines (apparently because many vortex lines no longer terminate on the grid after being ``shaken off'' by the emerging turbulent layer near the horizontal walls) while there is not much turbulence in
the central region yet.} 
\item{$\Omega t = 6$. The first peak has
disappeared in favor of the second one which got broadened --
at this stage the rectilinear vortices should have become
scrambled in the Ekman layers  near the top and bottom walls
while the ballistic charged vortex rings are still the dominant
transport of charge.} 
\item{$\Omega t = 12$. Now both fast peaks
have disappeared completely while the third broad peak has emerged
-- this means that a dense turbulent tangle has finally reached the
axial region.} 
\item{$\Omega t \rightarrow \infty$. The sharp
second peak has recovered but all others vanished -- after the
turbulence has decayed only ballistic charged vortex rings carry
the charge, neither the rectilinear array or turbulent tangle
contribute to the transport any more.} 
\end{enumerate}

\begin{figure}[h]
\begin{center}
\includegraphics[width=0.75\textwidth]{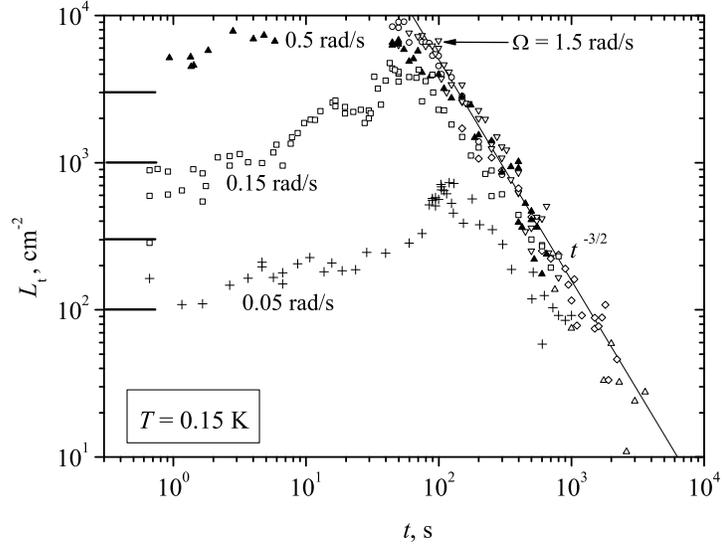}
\caption{$L_{\rm t}(t)$ at $T=0.15$\,K for four values of $\Omega$.
The average driving fields used for $\Omega = 1.5$\,rad/s: 5\,V/cm
($\circ$), 10\,V/cm ($\bigtriangledown$), 20\,V/cm ($\diamond$),
25\,V/cm ($\bigtriangleup$). At other values of $\Omega$ the
electric fields used were either 20\,V/cm (0.05\,rad/s) or 10\,V/cm
(0.5 and 0.15\,rad/s). The line shows the dependence
$t^{-3/2}$. The horizontal bars indicate the initial vortex
densities at steady rotation, $L=2\Omega/\kappa$, at
$\Omega=1.5$\,rad/s, 0.5\,rad/s, 0.15\,rad/s and 0.05\,rad/s (from top
to bottom).}
\label{MancRawSpinDownT}
\end{center}
\end{figure}

\begin{figure}[h]
\begin{center}
\includegraphics[width=0.75\textwidth]{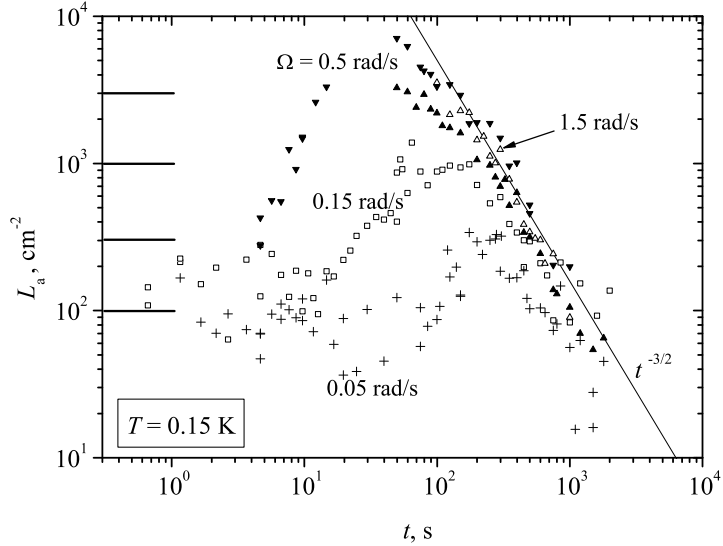}
\caption{$L_{\rm a}(t)$ at $T=0.15$\,K for four values of $\Omega$.
The average driving field used was 20\,V/cm in all cases except at
0.5\,rad/s where both 10\,V/cm ($\bigtriangledown$) and 20\,V/cm
($\bigtriangleup$) were used. The line shows the dependence
$t^{-3/2}$. The horizontal bars indicate the initial vortex
densities at steady rotation, $L=2\Omega/\kappa$, at
$\Omega=1.5$\,rad/s, 0.5\,rad/s, 0.15\,rad/s and 0.05\,rad/s (from top
to bottom).}
\label{MancRawSpinDownA}
\end{center}
\end{figure}

\begin{figure}[h]
\begin{center}
\includegraphics[width=0.75\textwidth]{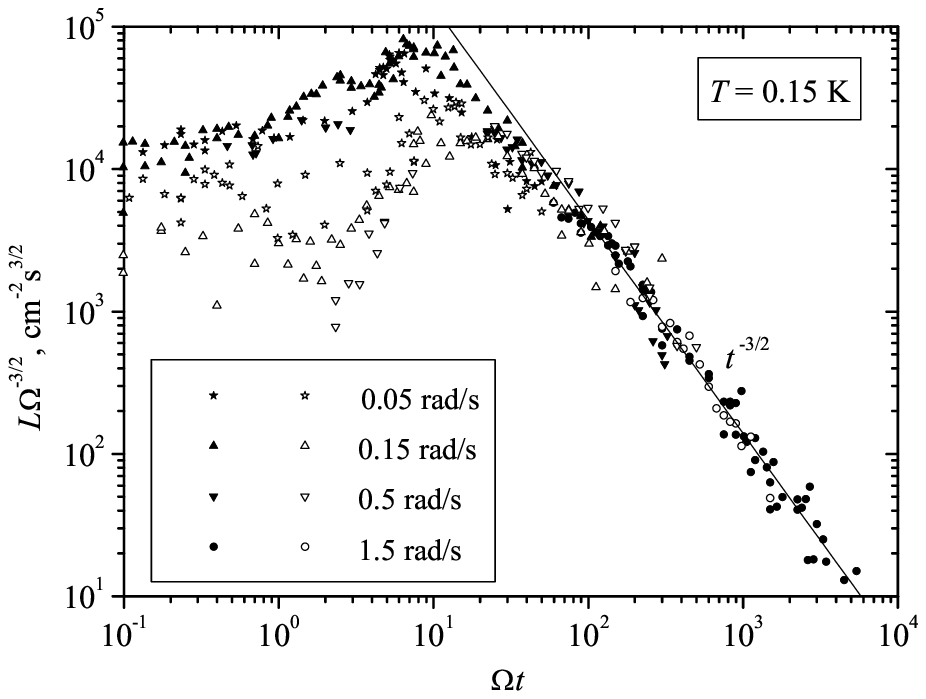}
\caption{$\Omega^{-3/2}L_{\rm t}(t)$ (filled symbols) and
$\Omega^{-3/2}L_{\rm a}(t)$ (open symbols) vs. $\Omega t$ for four
values of $\Omega$ at $T=0.15$~K. The straight line $\propto
t^{-3/2}$ guides the eye through the late-time decay.}
\label{MancSpinDownAT}
\end{center}
\end{figure}

In Fig.\,\ref{MancRawSpinDownT}, the measured densities of vortex
lines along the horizontal (transverse) direction, $L_{\rm t}$, are
shown for four different initial angular velocities $\Omega$.
During the transient, which lasts some $\sim 100\Omega^{-1}$,
$L_{\rm t} (t)$ goes through the maximum after which it decays
eventually reaching the universal late-time form of $L \propto
t^{-3/2}$. For $\Omega \geq 0.5$\,rad/s, the values of $L$ at
maximum were too high to be detected. The initial vortex densities
at steady rotation, $L=2\Omega/\kappa$, are shown by horizontal
lines. Similarly, the densities of vortex lines measured along the
vertical (axial) direction, $L_{\rm a}$, are shown in
Fig.\,\ref{MancRawSpinDownA}. 
To stress the scaling of the characteristic times
with the period of initial rotation $\Omega^{-1}$ and the universal
late-time decay $\propto t^{-3/2}$, we re-plot these data for different
$\Omega$ in Fig.\,\ref{MancSpinDownAT} rescaled accordingly.

We can see that at all $\Omega$
the transients are basically universal. Immediately after
deceleration, $L_{\rm t}$ increases, indicating the appearance of
the turbulent boundary layer at the perimeter, while $L_{\rm a}$
is stable at $L_{\rm i} \approx 2\Omega/\kappa$. Only at $\Omega t
\approx 3$, the latter starts to grow, signaling the destruction
of the rotating core with vertical rectilinear vortices. After
passing through a maximum at $\Omega t = 8$ and $\Omega t = 15$
respectively, $L_{\rm t}$ and $L_{\rm a}$ merge at $\Omega t \sim
30$ and then become indistinguishable. This implies that from now
on the tangle density $L$ is distributed nearly homogeneously. Eventually, after
$\Omega t \sim 100$, the decay takes its late-time form $L \propto
t^{-3/2}$ expected for quasiclassical isotropic turbulence, whose
energy is mainly concentrated in the largest eddies bound by the
container size $d$, but homogeneous on smaller length scales. We
hence assume that the turbulence in $^4$He at this stage is nearly
homogeneous and isotropic, and can apply Eq.\,(\ref{L-t-K}) to
extract the effective kinematic viscosity $\nu$ (see Discussion).

\begin{figure}[h]
\begin{center}
\includegraphics[width=0.75\textwidth]{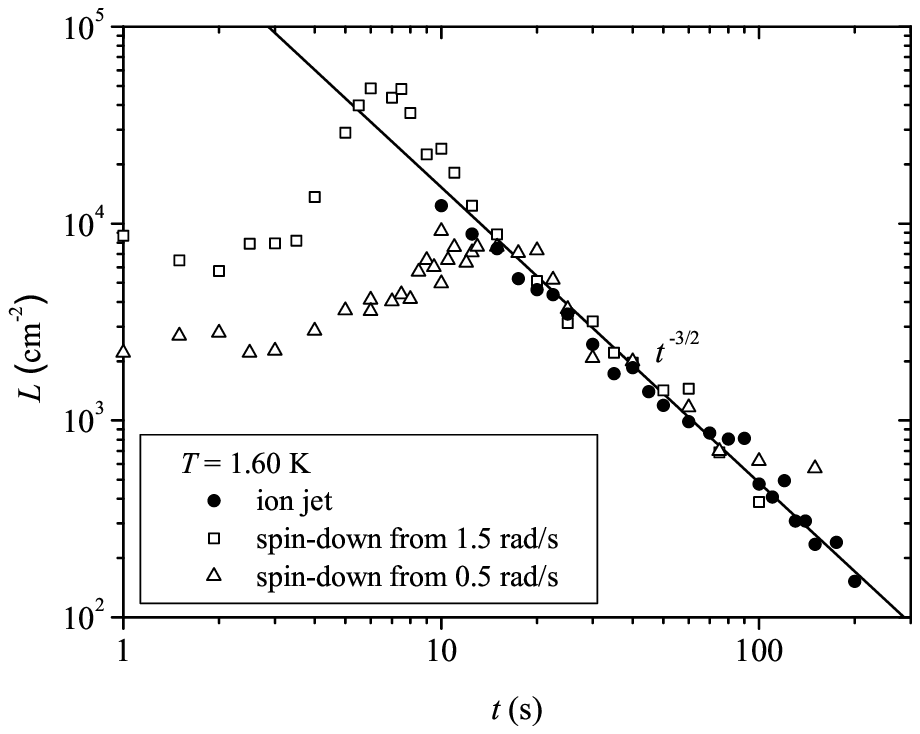}
\caption{Free decay of a tangle produced by an impulsive spin-down to rest 
\cite{Walmsley2007} from 1.5\,rad/s and 0.5\,rad/s, as well as by a jet of free ions
from the bottom injector ($\bullet$) \cite{Walmsley2008}, at $T=1.60$\,K.
All tangles were probed by pulses of free ions in the horizontal
direction. The line $L\propto t^{-3/2}$ corresponds to
Eq.\,(\ref{L-t-K}) with $\nu = 0.2\kappa$.}
\label{MancIonJetHT}
\end{center}
\end{figure}

\begin{figure}[h]
\begin{center}
\includegraphics[width=0.75\textwidth]{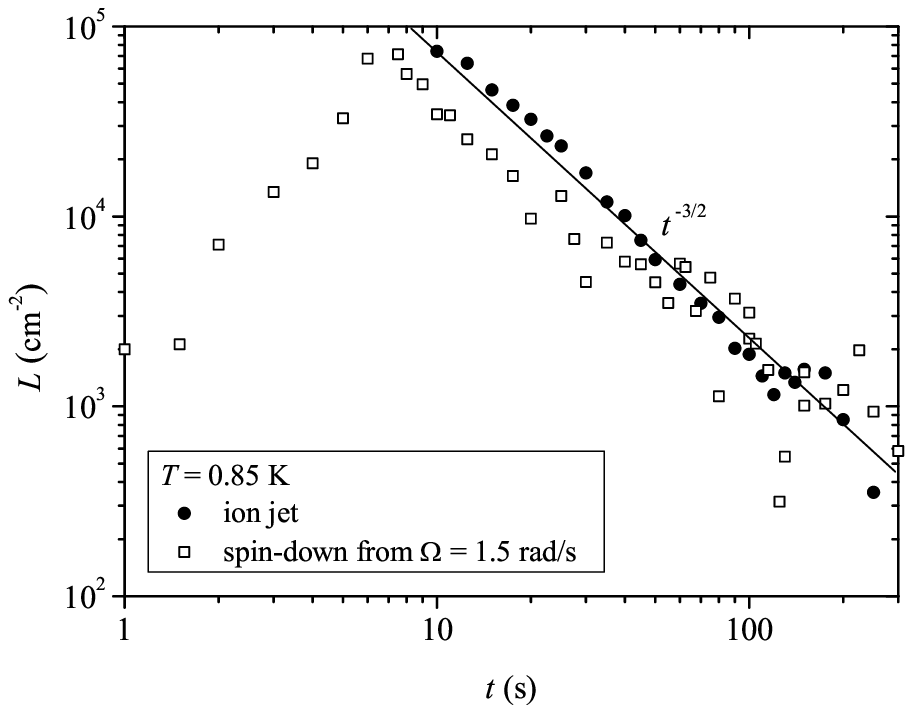}
\caption{Free decay of a tangle produced by an impulsive spin-down to rest \cite{Walmsley2007} from
1.5\,rad/s, as well as by a jet of free ions
from the bottom injector ($\bullet$) \cite{Walmsley2008}, at $T=0.85$\,K. All tangles were probed by pulses of
free ions in the horizontal direction. The ion jet data are the
average of nine measurements at each particular time but the
spin-down data show individual measurements.}
\label{MancIonJetLT}
\end{center}
\end{figure}

At all temperatures the transients $L_{\rm t}(t)$ after a
spin-down are, in first approximation, universal, \textit{i.e.}
the timing of the maximum is the same $\Omega  t\approx 7$ and the
amplitudes of the maximum are comparable. This supports our
approach to turbulent superfluid helium at large length scales as
to an inviscid classical liquid, agitated at large scale and
carrying an inertial cascade down the lengthscales, independent of
temperature. On the other hand, as the temperature rises and
the damping through the normal component increases, changes in the
shape of transients might be expected. Indeed, one can see that
the  slope of $L(t)$ after the maximum but before reaching the
ultimate late-time decay $L\propto t^{-3/2}$ (\textit{i.e.} for
$10 < \Omega t < 100$) is changing gradually with increasing
temperature from being less steep than $t^{-3/2}$ at $T=0.15$\,K to
more steep than $t^{-3/2}$ at $T=1.6$\,K
(Fig.\,\ref{MancIonJetHT}). At temperatures around $T
=0.85$\,K (Fig.\,\ref{MancIonJetLT}) it nearly matches $t^{-3/2}$,
thus making an erroneous determination of $\nu$ possible by taking
this part of the transient for the late-time decay $L \propto
t^{-3/2}$. Indeed, in the first publication \cite{Walmsley2007},
parts of some transient at $T=0.8-1.0$\,K for as early as $\Omega t
> 15$ were occasionally used to be fitted by $L \propto t^{-3/2}$
that often resulted in overestimation of the value for $\nu$. To
rectify this, we have re-fitted the datasets used in the original
publication as well as subsequent measurements with $L\propto
t^{-3/2}$ for spin-downs using the following rules: for
$T\leq0.5$\,K only points for $\Omega t > 300$ were used, between
0.5\,K and 1.0\,K only points for $\Omega t>150$ used and at
$T>1.0$\,K only $\Omega t > 75$ were used. This resulted in slight
systematic reduction of the extracted values of $\nu(T)$ at
temperatures 0.8--1.2\,K from those published in
\cite{Walmsley2007}; what looked as a rather steep drop in
$\nu(T)$ now occurs at 0.85 -- 0.90\,K and is somewhat reduced in magnitude.

\begin{figure}[h]
\begin{center}
\includegraphics[width=0.75\textwidth]{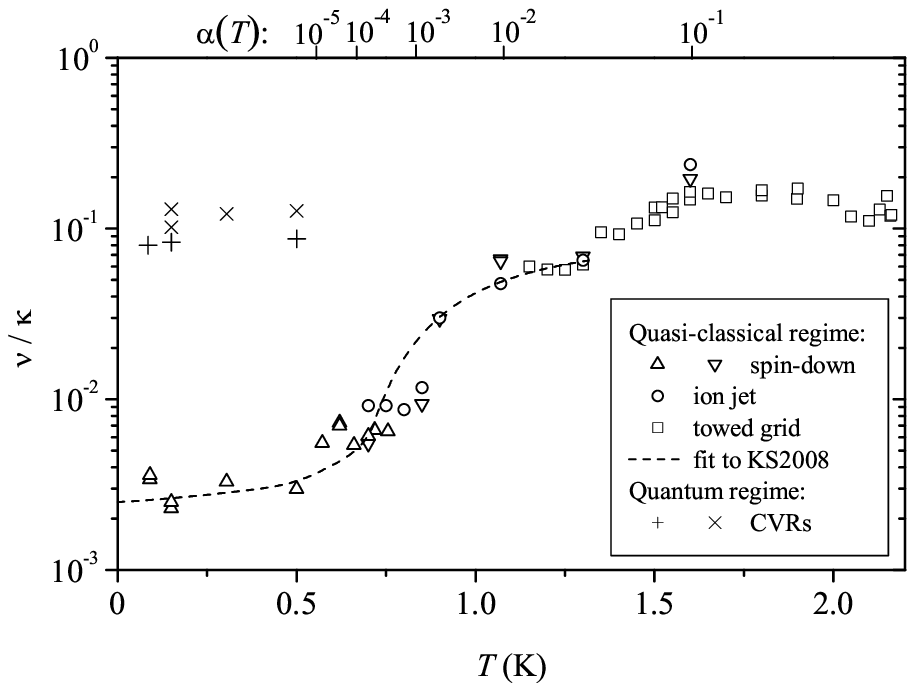}
\caption{The effective kinematic viscosity $\nu(T)$ for quasiclassical turbulence in $^4$He: $\nu(T)$ infered using Eq.\,(\ref{L-t-K}) from the free decay $L\propto t^{-3/2}$ of tangles 
produced by impulsive spin-down (open triangles
\cite{Walmsley2007}), ion jet (open circles
\cite{Walmsley2008}) and towed grid (open
squares \cite{Stalp2002}). The dashed curve is a fit
to our data (open triangles) by theory 
\cite{KS2008}. For comparison, the values for $\nu(T)$ for the quantum regime of turbulence in $^4$He, generated by a beam of colliding CVRs, are shown by crosses \cite{Walmsley2008}.}
\label{MancNuPrime} 
\end{center} 
\end{figure}

\section{Discussion}


Let us begin from commenting on the significance of observation of $L\propto t^{-3/2}$ decay in a broad range of decay time $t$. For quasi-classical turbulence, at lengthscales approaching the cross-over length $\sim \ell$, below which the classical cascade no longer holds, there might exist an additional ``equipartitioned'' component of vorticity that does not contribute to the steady-state energy cascade \cite{LNR2007}. Also, in any type of quantum turbulence, unlike the classical case, there exists an additional {\it quantum} contribution to the total energy resulting from the quantized flow around individual filaments, $E_{\rm q} = B L$ where $B \approx const$ \cite{Walmsley2008}. During the free decay, both these reservoirs release additional energy to be dissipated. For the math leading to Eq.\,(\ref{L-t-K}) to hold, these energy fluxes should not add to the flux $\epsilon$ of classical energy in Eq.\,(\ref{EDot}) (or should at least have the same time-dependence $\propto t^{-3}$ as the classical one, which is unlikely). At early stages of the decay $|\dot{E_{\rm q}}|$ is always small compared to $\epsilon$, but estimates show that it might become comparable at very late times, especially in the $T=0$ limit. The fact that we never observed deviations from the $L\propto t^{-3/2}$ decay at late times suggests that these fluxes are simply not coupled to Eq.\,(\ref{EDot}). For instance, this would be the case if this equation describes some $L$-dependent limitation for the energy flux approaching the classical-quantum crossover scales $\sim \ell$ as the quantum flux $|\dot{E_{\rm q}}|$ can only be injected at shorter lengthscales $< \ell$.

In Fig.\,\ref{MancNuPrime} we plot the values of $\nu(T)$ as extracted from the late-time decays $L \propto t^{-3/2}$ using Eq.\,(\ref{L-t-K}). We assumed that, at all temperatures, $C=1.5$ and $k_1=2\pi/d$, and the only temperature-dependent parameter in Eq.\,(\ref{L-t-K}) is $\nu(T)$. In principle, the interaction between the vortex tangle and container walls changes with temperature (at high temperatures, via mutual friction with the normal component which is viscously coupled to walls, but  at low temperatures, only through the non-axially-symmetric shape of the container, including both its cubic shape and surface roughness). Hence, the boundary conditions of the quasi-classical turbulent flow can change with temperature, thus affecting the numerical prefactor in the relation for the cut-off wavenumber for the largest eddies,  $k_1=2\pi/d$. Also, at low temperatures, the hard core of the giant vortex might survive for longer after the spin-down. This would result in an anisotropic rotating turbulence whose energy-containing eddy (the giant vortex) might release its energy slower than the classical isotropic 3d turbulence: as a result the effective value of the constant $C$ might decrease with decreasing temperature. The fact that we have shown experimentally that the vortex density $L$ is distributed homogeneously throughout the cell at late times does not guarantee that no such large-scale anisotropy exists. Also, there is no proof that the coarse-grained hydrodynamic equations have the same strength of the non-linearity as the classical Navier-Stokes equation; hence, the effective Kolmogorov constant can in general differ from the classical value (and can depend on temperature too). While we believe that these changes with temperature of the effective parameters $C$ and $k_1$ might be unlikely, we cannot rule them out at this stage. Obviously, further spin-down experiments in containers of different shapes as well as with the turbulence generated by other means (e.g. rotating agitators \cite{Maurer1998}, towed grid \cite{Stalp1999,Niemela2005}, vibrating grid \cite{Bradley2006}, propagating turbulent front \cite{Eltsov2007}, counterflow \cite{Chagovets2008}, ion jet \cite{Walmsley2008}) and numerical simulations of the tangle in a cubic cell upon a spin-down can clarify these issues. For example, the fact that quasi-classical tangles, generated by ion jets in a non-rotating cryostat \cite{Walmsley2008}, yield the same values of $L(t)\propto t^{-3/2}$ (see Figs.\,\ref{MancIonJetHT}--\ref{MancNuPrime}), and hence of $\nu(T)$, down to $T=0.7$\,K as those after spin-down strongly suggests that slowly decaying background rotation surviving long time after a spin-down, if any, might be irrelevant. 

For comparison, in Fig.\,\ref{MancNuPrime}, the high-temperature data for quasi-classical turbulences generated by towed grids and counterflow in channels with square cross-section are shown too. Generally they are in good agreement with our results. We also show the low-temperature values of $\nu \approx 0.1\kappa$ for quantum tangles generated by short injections of CVRs, that apparently have little large-scale flow. The main observation is that the infrerred values of $\nu(T)$ for quasi-classical tangles show a substantial decrease with cooling below $T\sim 1$\,K, eventually reaching the temperature-independent plateau of $\nu \approx 0.003\kappa$ that is much smaller than that for the quantum tangles. 

At the moment there exist two different theoretical models explaining the dropping of quasiclassical $\nu(T)$ approaching the $T=0$ limit. The ``bottleneck'' scenario by L'vov, Nazarenko and Rudenko \cite{LNR2007} explains the pile-up of excess vorticity on the classical side of the cross-over lengthscale $\ell$ as a result of poor matching of the kinetic times of the classical and Kelvin-wave cascades; this pile-up can only happen at sufficiently low temperatures when the cascading energy can reach the Kelvin-wave cascade. On the other hand, Kozik and Svistunov \cite{KS2008} explain the enhancement of the total length $L$ at low temperatures as a result of extra contribution at lengthscales shorter than $\ell$ caused by the fractalization of the tangle after various reconnections processes become possible at low temperatures where Kelvin waves are hugely underdamped. It seems, at this stage we cannot discriminate conclusively between the two alternative explanations; it is possible that both mechanisms contribute. Future experiments as well as numerical and analytical calculations should help clarify the issue. For example, it is important to know the rate of reconnections as function of the net polarization of the tangle. We are currently conducting experiments on the dynamics of  tangles at continuous rotation that might advance our understanding. 

To conclude, we would like to note that these are the first
measurements on superfluid turbulence below 1\,K after fifty years of
research which now point out (i) how to create homogeneous isotropic 
turbulence (at least in a transient state) and (ii) how to measure it. The
road is now open to study the remaining open questions in the $T=0$ limit,
starting from a convincing experimental proof of equation \ref{EDot} on which
the analysis in the paper is based.

\vspace{0.3cm}\noindent {\bf Acknowledgements} We thank Steve May, Stan Gillott and Mark Sellers for their contributions to the construction and maintenance of the apparatus, as well as the referee who suggested to include the concluding paragraph. Support was provided by EPSRC under GR/R94855 and EP/E001009. 

\end{document}